\documentclass[%
reprint,
superscriptaddress,
amsmath,amssymb,
aps,
floatfix,
]{revtex4-2}

\usepackage{graphicx}
\usepackage{dcolumn}
\usepackage{bm}
\usepackage{siunitx}
\usepackage{braket}
\usepackage{bm}
\usepackage{color}
\usepackage[pdfencoding=auto, psdextra]{hyperref}
\usepackage{cleveref}
\Crefformat{figure}{#2Fig.~#1#3}
\Crefmultiformat{figure}{Figs.~#2#1#3}{ and~#2#1#3}{, #2#1#3}{ and~#2#1#3}
\usepackage{float}
\usepackage[flushleft]{threeparttable}
\usepackage[ansinew,utf8]{inputenc}
\usepackage{booktabs}
\usepackage{xr}
\makeatletter
\newcommand*{\addFileDependency}[1]{
  \typeout{(#1)}
  \@addtofilelist{#1}
  \IfFileExists{#1}{}{\typeout{No file #1.}}
}
\makeatother

\newcommand*{\myexternaldocument}[1]{
    \externaldocument{#1}
    \addFileDependency{#1.tex}
    \addFileDependency{#1.aux}
}
\myexternaldocument{supplementary}

\newcolumntype{_}{>{\global\let\currentrowstyle\relax}}
\newcolumntype{^}{>{\currentrowstyle}}

\newcommand{\bullets}{0} 
\begin{document}
	
	\preprint{APS/123-QED}
	
	\title{Vanadium in Silicon Carbide: Telecom-ready spin centres with long relaxation lifetimes and hyperfine-resolved optical transitions}

    \author{Thomas Astner}
    \affiliation{Vienna Center for Quantum Science and Technology, Universit\"at Wien, Boltzmanngasse 5, 1090 Vienna, Austria}
    \author{Philipp Koller}
    \affiliation{Vienna Center for Quantum Science and Technology, Universit\"at Wien, Boltzmanngasse 5, 1090 Vienna, Austria}
	\author{Carmem M. Gilardoni}
	\affiliation{Zernike Institute for Advanced Materials, University of Groningen, NL-9747AG Groningen, The Netherlands}
    \author{Joop Hendriks}
	\affiliation{Zernike Institute for Advanced Materials, University of Groningen, NL-9747AG Groningen, The Netherlands}
    \author{Nguyen Tien Son}
    \author{Ivan G. Ivanov}
    \author{Jawad Ul Hassan}
	\affiliation{Department of Physics, Chemistry and Biology, Link\"oping University, SE-581 83 Link\"oping, Sweden}
    \author{Caspar H. van der Wal}
	\affiliation{Zernike Institute for Advanced Materials, University of Groningen, NL-9747AG Groningen, The Netherlands}
    \author{Michael Trupke}
    \affiliation{Vienna Center for Quantum Science and Technology,  Universit\"at Wien, Boltzmanngasse 5, 1090 Vienna, Austria}
    
	\date{\today}
	
	\begin{abstract}
		Vanadium in silicon carbide (SiC) is emerging as an important candidate system for quantum technology due to its optical transitions in the telecom wavelength range. However, several key characteristics of this defect family including their spin relaxation lifetime ($T_1$), charge state dynamics, and level structure are not fully understood. In this work, we determine the $T_1$ of an ensemble of vanadium defects, demonstrating that it can be greatly enhanced at low temperature. We observe a large spin contrast exceeding $90\,\%$ and long spin-relaxation times of up to $\SI{25}{\second}$ at \SI{100}{\milli\kelvin}, and of order $\SI{1}{\second}$ at $\SI{1.3}{\kelvin}$. These measurements are complemented by a characterization of the ensemble charge state dynamics. The stable electron spin furthermore enables high-resolution characterization of the systems' hyperfine level structure via two-photon magneto-spectroscopy. The acquired insights point towards high-performance spin-photon interfaces based on vanadium in SiC. 
	\end{abstract}
	
	\maketitle
	
	\section{Introduction}
	\label{sec:intro}
		Semiconductor materials such as silicon, diamond, and silicon carbide (SiC) are excellent hosts for defect centers given their large band gaps, hardness, and quiescent magnetic environment \cite{Weber2010,Wolfowicz2021,Chatterjee2021}. Numerous defects in these materials present transitions in the optical domain, coupled with long spin relaxation times \cite{Doherty2013,Christle2015,Widmann2015,Fuchs2015,Simin2016,Morse2017,Bosma2018,Beaufils2018}. These features make them attractive for applications in quantum sensing, computation, and communication \cite{Atature2018}. However, most of these defects have optical transitions in the visible range, posing challenges for low-loss photonic systems and long-distance communication \cite{Janitz2020,Lukin2020}.
    
    Optically active spin centers in SiC, a semiconductor widely used in the microelectronics industry, have gained widespread interest in the community since SiC hosts several emitters which may be suitable for quantum communication and computing applications in the telecom wavelength range \cite{Babunts2000,Magnusson2005,Koehl2011,Zargaleh2016}.
    
    In this work, we study the properties of neutral vanadium ($\mathrm{V}^{4+}$) defect centers in the 4H and 6H polytypes of SiC between \SI{100}{\milli\kelvin} and \SI{3}{\kelvin}. \cite{Kunzer1993,Reinke1993,Schneider1990}. This transition metal defect offers bright and fast optical transitions in the telecom O-band around \SI{1.3}{\micro\meter}, and is therefore highly suited for spin-photon entanglement and integration into ultralow-loss enhancement structures for long-range fiber based communication in silica optical fibers \cite{Spindlberger2019,Wolfowicz2020,Vasconcelos2020,Lukin2020,Fait2021}. 
	The family of vanadium emitters has been investigated extensively as it is a common impurity, and has been used for decades in charge compensation for industrial wafer-grade SiC \cite{Schneider1990,Jenny1995}. However, many of the most important optical and spin properties for quantum applications remain to be determined.
	
	Here, we investigate several of these aspects in depth by performing all-optical spectroscopy, charge depletion, and spin relaxation measurements in the temperature regime below \SI{2.5}{\kelvin}.
	Finally, two-laser spectroscopy of the defects enables detailed insight into the rich hyperfine level structure.
	These measurements lay the groundwork for methods enabling precise initialization and control of the electronic and nuclear spin degrees of freedom.

	\begin{figure*}[t]
	\centering
	\includegraphics[scale=0.9,page=1]{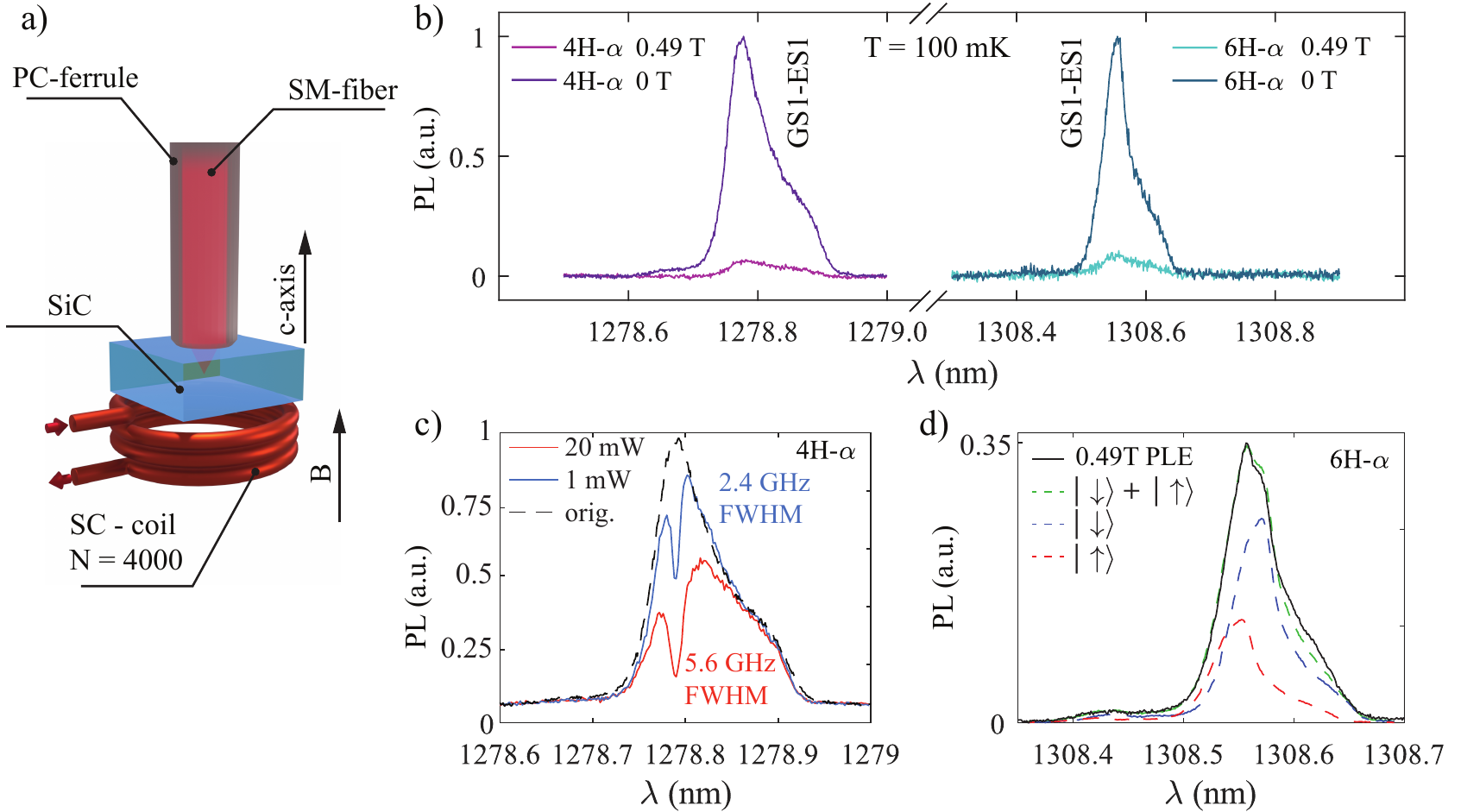}
	\caption{\textbf{Optical spectroscopy of V$^{4+}$ ensembles in 4H- and 6H-SiC.}
	\textbf{a)} Experimental set-up: A SiC sample is mounted at the lowest stage of a dilution refrigerator operating at \SI{100}{\milli\kelvin}.
	A superconducting solenoid with 4000 turns is located beneath the sample, and is used to generate DC magnetic fields of up to \SI{0.49}{\tesla}, parallel to the c-axis.
	The optical interface with the vanadium ensemble is a spring-loaded, commercial flat-polished (PC) fiber ferrule containing a single-mode fiber (SMF-28), positioned on the top surface of the sample and on axis with the solenoid.
	Resonant infrared and green (\SI{520}{\nano\meter}) laser light is transmitted through the fiber and illuminates the vanadium defects in the SiC sample.
	\textbf{b)} Resonant PL spectroscopy with and without magnetic field for the $ \alpha $ - sites in 4H- and 6H-SiC.
	By tuning the laser through resonance we observe bright emission in the phonon sideband.
	Setting the bias magnetic field to \SI{0.49}{\tesla} results in a low amplitude in the spectroscopy signal. This effect is attributed to optical pumping of spin population into the opposite spin state, and thus out of resonance.
	The data traces are peak-normalized for clarity. The traces were collected without green excitation.
	\textbf{c)} Charge holeburning. Driving the optical GS1-ES1 transition in 4H-$\alpha$ continuously results in a two-photon process that ionizes the defect centres.
	The PL signal drops over time and a spectral hole is formed at the wavelength of the driving laser. 
	The hole width and depth are both dependent on the applied drive power.
	After bleaching, the hole persists over several hours at \SI{100}{\milli\kelvin}. 
	The power given in the figure refers to the laser output.
	\textbf{d)} PL signal under continuous above band-gap illumination with a green laser at  \SI{0}{\tesla} and \SI{0.49}{\tesla} magnetic field.
	The spectrum obtained with magnetic field is fitted with a sum of two copies of the zero-field lineshape with different amplitudes and frequency offsets, accounting for the two symmetrically Zeeman-shifted spin states and their relative populations (see text).
	}
	\label{fig:PLEspec}
\end{figure*}	
	
	\section{Optical spectroscopy of $^{51}$V ensembles}
	\label{sec:opticalspec}
        The $^{51}$V isotope substitutes a silicon site as a dopant in the SiC lattice.
        Depending on the polytype of SiC, the vanadium atom can occupy inequivalent lattice sites.
        Here we study vanadium in the stable charge state $\mathrm{V^{4+}}$ (neutral) in 4H- and 6H-SiC.              
        We focus on sites that have an optical transition close to \SI{1.3}{\micro\meter}, denoted by the site name $\alpha$.
        
        The electronic level structure of $\mathrm{V^{4+}}$ is comparable to that of the group IV defects in diamond, such as the neutral silicon vacancy \cite{Thiering2020,Hepp2014}.
        V$^{4+}$ has a single unpaired electron that is coupled to the nuclear spin of $^{51}$V. The center has two Kramers doublet ground states  (GS1, GS2) and two optical excited (ES1, ES2) \cite{Wolfowicz2020}.

        The spin properties of these states can be described by a Hamiltonian in the following form, that includes Zeeman and hyperfine interaction terms:
        \begin{equation}
            H=\mu_{\mathrm{B}} \boldsymbol{B}_{0} \cdot \boldsymbol{g} \cdot \boldsymbol{S}-\mu_{\mathrm{N}} g_{\mathrm{N}} \boldsymbol{B}_{0} \cdot \boldsymbol{I}+\boldsymbol{S} \cdot \boldsymbol{A} \cdot \boldsymbol{I}.
            \label{eq:sysham0}
        \end{equation}
        Here $ \bm{S} $ and $\bm{I} $ are the electronic ($S=1/2$) and nuclear ($I=7/2$) spin operators, $ g_N $ and $ \bm{g} $ are nuclear and effective electron g-factors, $ \bm{B_0} $ a static magnetic field and $ \bm{A} $ the hyperfine interaction tensor between the nuclear and effective electron spins.
        Both ground and excited states can be described with this Hamiltonian, but differ in their hyperfine interaction tensor and electron g-factor.
        In \Cref{tab:table} the parameters of the Hamiltonian for the $\alpha$ sites in both polytypes are displayed.
        
        We first discuss resonant photoluminescence excitation (PLE) measurements of the ensemble optical transition, performed using excitation with a tunable laser and collection of the photoluminescence emission from the subsequent decay.
        Stray counts from the excitation light are strongly suppressed by collecting only photons from the phonon sideband of the emission.
        We interface a vanadium ensemble in commercial grade wafers of 4H-SiC or 6H-SiC with a single mode fiber (SMF-28) with a PC-ferrule end. The ferrule is spring loaded and aligned parallel to the wafer c-axis. Light reaching the sample is attenuated by $26\,$dB-$29\,$dB on its path from the tunable laser.
        Additionally, a superconducting solenoid is placed below the sample, to apply magnetic bias fields of up to \SI{0.49}{\tesla}  along the c-axis. 
        A schematic representation of the setup is shown in \Cref{fig:PLEspec} a).
        
        In our PLE spectroscopy data, recorded at a temperature of \SI{100}{\milli\kelvin} and plotted in \Cref{fig:PLEspec} b), we identify the GS1-ES1 transitions of the $\alpha$ sites in 4H- and 6H-SiC.
        The wavelengths of the transitions can be found in \Cref{tab:table} and are in agreement with previous studies \cite{Kunzer1993,Wolfowicz2020}. The lineshape and linewidth are strongly dependent on the crystal properties. Epitaxially grown samples \cite{Karhu2019} present a strongly asymmetric lineshape which is markedly narrower (see supplementary section \ref{suplsec_PLE}). 
        Transitions involving the orbital states GS2 and ES2 were not observed.
        We attribute their absence to the thermal depopulation of the second ground state GS2 \cite{Kunzer1993,Spindlberger2019,Wolfowicz2020} and the orientation of the resonant illumination and collection of photons (the k-vector of the excitation light is parallel to the c-axis) \cite{Kunzer1993,kaufmann1997}.
        Indeed, in accord with the crystal-field model proposed in Ref. \cite{kaufmann1997}, transitions from the lowest in energy counterpart of the GS (GS1) couple to ES1 via photons polarized perpendicular to the c-axis ($ E \bot \mathrm{c} $), while the transition GS1 - ES2 requires $ E \parallel \mathrm{c} $ polarization of the incident light, which is not available within our experimental geometry.
        When a bias magnetic field of \SI{0.49}{\tesla} is applied, the luminescence decreases significantly. 
        In order to understand this behaviour, and to distinguish it from ionization effects, two further sets of PLE measurements were performed.  
        
               \begin{table}
    \caption{System parameters.}
    \label{tab:table}
    \centering
    \begin{threeparttable}
            \begin{tabular}{@{}lll@{}}
            \toprule
            Name                           & 4H-$\alpha$             & 6H-$\alpha$             \\ \midrule
            ES1-GS1                        & \SI{1278.78}{\nano\meter}    & \SI{1308.56}{\nano\meter}   \\
            GS1: $\mathrm{g_{xx,yy}, g_{zz}}$           & 0\tnote{\textasteriskcentered}, 1.748\tnote{\textasteriskcentered}     & 0\tnote{\textasteriskcentered}, 1.749\tnote{\textasteriskcentered}     \\
            GS1: $\mathrm{A_{xx}, A_{yy}, A_{zz}}$ (\si{\mega\hertz}) & 165\tnote{\textasteriskcentered},-165\tnote{\textasteriskcentered},232(5)\tnote{\textasteriskcentered} & 165\tnote{\textasteriskcentered},-165\tnote{\textasteriskcentered},232(5)\tnote{\textasteriskcentered} \\
            ES1: $\mathrm{g_{xx,yy}, g_{zz}}$           & -, \SI{2.18\pm0.01}{}        & -,2.24\tnote{\textasteriskcentered}         \\
            ES1: $\mathrm{A_{xz}, A_{zz}}$ (\si{\mega\hertz})     & 75(4), -213(4)  & 20,200(20)     \\ \bottomrule
            \end{tabular}
      \begin{tablenotes}
        \footnotesize
        \item[\textasteriskcentered] Values taken from literature \cite{Baur1997,Wolfowicz2020}.
      \end{tablenotes}
    \end{threeparttable}
  \end{table}
        \textit{Charge holeburning:}  In high-power PLE measurements at \SI{0}{\tesla}, we observed a decay of the fluorescence signal in 4H-$\alpha$ after prolonged exposure to resonant excitation. 
        A bleaching experiment was therefore performed by tuning the excitation laser to the peak of the resonance and illuminating it for \SI{60}{\second} with different laser output powers.
        As can be seen in \Cref{fig:PLEspec} c), a spectral hole at the laser frequency is formed.
        This hole persists for several hours and is due to ionization of resonant defects \cite{Kummer1997}.
        Both the width and the depth of the spectral hole increase with laser power, resulting in spectral holes with a width of \SIrange[]{2}{6}{\giga\hertz} and a reduction in fluorescence by up to 80\%. The width is far greater than the laser linewidth, which has a \SI{400}{\kilo\hertz} typical FWHM (full width at half maximum).
        Detailed information and plots of the hole-burning experiments can be found in the supplementary section \ref{suplsec_hole}.
        Adding green illumination (\SI{520}{\nano\meter}) to the resonant excitation not only prevents bleaching, but increases the observed fluorescence count rate.
       In the experiment, we use this mechanism to restore the original spectrum without hole by reviving the ionized charges.
       Both the photo-bleaching in 4H and the absence of this effect in 6H has been observed and is in agreement with previous work using this material \cite{Wolfowicz2020}.
       However, we assume that the charge stability is highly dependent on the crystal quality and dopant concentration as, contrary to our measurements, photo-bleaching has also been observed in 6H \cite{Kummer1997}.
       These observations provide a basis for further investigations towards a complete understanding of the charge state dynamics of vanadium in 4H- and 6H-SiC. We underline that, even at the highest input power of \SI{20}{\milli\watt}, charge bleaching occurs on a timescale of order \SI{10}{\second}.
       
        \textit{Spin depletion:} Next, we discuss the PLE spectra of both defects with an applied magnetic field of \SI{0.49}{ \tesla}, without green illumination.
        As can be seen in \Cref{fig:PLEspec} b), application of a bias field results in a strong reduction of the fluorescence when compared to the zero field measurement.
        In contrast to the charge holeburning measurements, each frequency was only illuminated for \SI{0.5}{\second} at a laser output power of -6dBm (\SI{250}{\micro\watt}), underlining that a different mechanism than charge depletion is at play.
        The different g-factors of ES1 and GS1 lead to a splitting between the optical transitions depending on the electron spin state due to the Zeeman effect.
        During the PLE sweep of the laser at high magnetic fields, the electrons can decay into the opposite spin state that is not resonant with the laser and thus no longer contribute to the fluorescence signal.
        This effect can be observed in both 4H-$\alpha$ and 6H-$\alpha$, and leads to a spin population contrast in excess of 90\%.
        
        Adding (low power) green illumination to the PLE measurement at a high magnetic field leads to a partial recovery of the fluorescence.
        This observation indicates that green illumination excites a wide range of transitions, promoting electrons to the conduction band and ionizing surrounding charge traps. These disruptive processes are likely to lead to the observed mixing of the spin states.
        The measured spectrum is therefore a result of spin mixing by the low power green illumination and simultaneous spin pumping by the resonant laser during the PLE sweep.
        The agreement of the measured spectrum with a fit using two copies of the lineshape obtained at zero field, each with different amplitudes, shifted symmetrically to account for the Zeeman effect, supports this interpretation. The observed imbalance between spin states corresponds to an effective spin temperature of $\sim$\SI{230}{\milli\kelvin}, significantly higher than the cryostat temperature during this scan (\SI{170}{\milli\kelvin}). The green laser can therefore be used to initialize the spins in a mixed state. We use this mechanism for the spin lifetime measurements in the following section.

\ifnum \bullets>0 
    \begin{itemize}
        \item \textbf{Optical Spectroscopy of V ensembles}
        \item What is V in Sic: Sits at silicon site of SiC  

        \item Different polytypes of Sic - 4H and 6H
        \item different lattice sites for V
        \item V alpha sites in 4H and 6H SiC are investigated
        \item V behaves like an effective electron spin 1/2 system with two ground and excited states 
        \item V has a nuclear spin of 7/2
        \item system can be described with an Hamiltonian in the form... \cref{eq:sysham}
        \item All mentioned states (GS and ES) can have different g-factors
        \item Literature value g-GS1 of 4H/6H alpha: Kaufmann1997, J.Baur 1997
        \item Study defect we use resonant PLE with detection in phonon sideband.
        \item setup like shown in figure 1 a)
	    \item \textbf{Figure 1 a)}
	    \item PLE spectroscopy with SM fiber on SiC chip, excitation and bias magnetic field along c-axis, green and IR via same fiber
	    \item Bias magnetic field of up to 480mT at 2A current and 4k windings, solenoid placed below the SiC chip
	    \item \textbf{Figure 1 b)}
	    \item Resonant excitation with IR laser, detection of phonon sideband emission
	    \item Features at 100mK temperature, identification of alpha sites in 4H and 6H - their GS1-ES1 transition
	    \item No GS2-ES2 transition are observed in this temperature regime - no population in GS2 - see spindelberger, awschalom
	    \item No GS1-ES2 transitions are visible because of excitation and collection along c-axis (k-vector of light parallel to c-axis) \cite{Kunzer1993}
        \item \textbf{Figure 1 c)}
        \item PLE shows charge bleacing in 4H-alpha - investigation: charge hole burning in 4H - not in 6H
        \item Charge state burning already observed in 4H and 6H SiC  cite: Kummer 1997, Awschalom
        \item Ionization possible with strong IR cw laser (60 sec pump) with different laser output power
        \item Lifetime of a charge hole is more than 2 hours
        \item Hole width and depth is power dependent with constant pump time
        \item Charge state switching, distinguish between spin pumping and charge state switching, power dependence - orders of magnitude difference in power between spin pumping and charge state switching
        \item charge hole lifetime over hours, spin hole much shorter
        \item charge holes can be removed with green laser - reset to original lineshape
	    \item charge hole sizes?
	    
	    \item \textbf{Again figure b)}
	    \item Comparing PLE with and without bias field, features vanish with field and no above bandgap 
	    \item spin pumping in other state is possible because g-factor difference of GS1 and ES1
	    \item GS1 spins split, ES1 splits more, g-factor difference as above, allows for optical splitting
	    \item spec sweep already pumps population in other state, pumping contrast of 90 percent

	    \item \textbf{Figure 1 d)}
	    \item PLE with cw green illumination with and without bias magnetic field
	    \item analysis of feature with an without magnetic field, fitting the spectrum with the two lineshapes obtained from the zero field feature
	    \item With green line recovers, but lower signal amplitude, more is pumped by IR than recovered by green
	    \item fitting the line shape with two indidivudual lineshapes - numerical line shape from 0mT data
	    \item fit parameters are amplitude and splitting, both spin populations are visible with green on, raising spin temperature (not same amplitude, not enough green)
	\end{itemize}
\fi	
		\label{sec:lifetime}
		\begin{figure*} [t]
		\centering
		\includegraphics[scale=0.9,page=2]{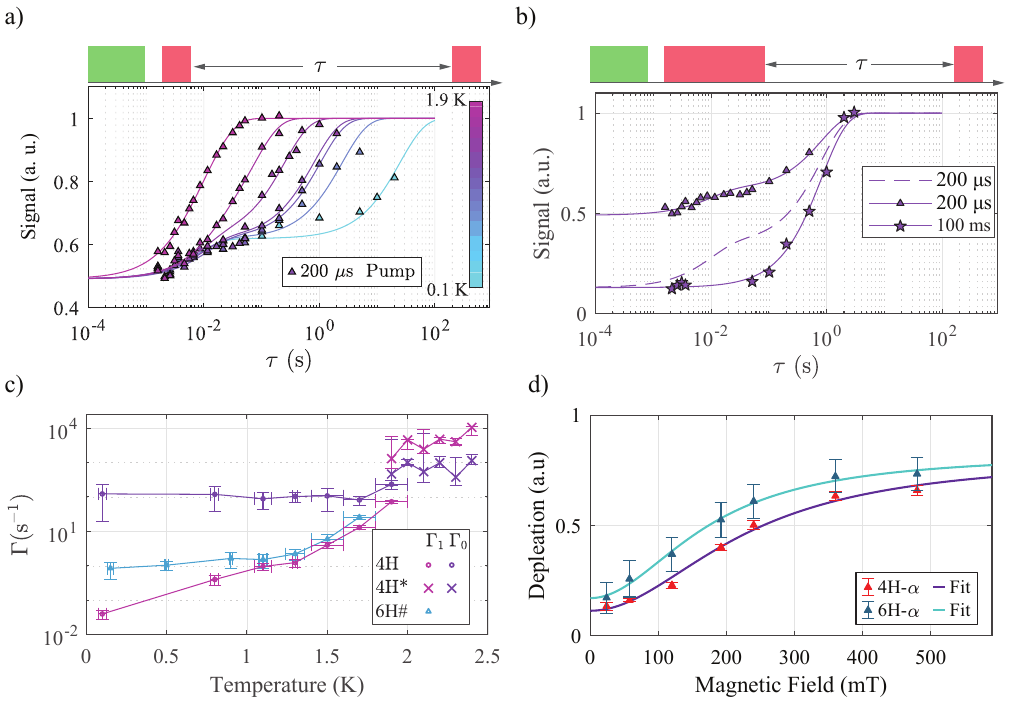}
		\caption{\textbf{Spin lifetime of vanadium in 4H and 6H SiC.}
		    \textbf{a)} Recovery signal obtained from a spin depletion and recovery sequence, shown above the plot, on 4H-$\alpha$ at \SI{0.49}{\tesla}.
		    First, the system is initialized in a mixed state with a \SI{0.5}{\milli\second} long green laser pulse (green rectangle), followed by pulse of $\SI{200}{\micro\second}$ length, on resonance with the GS1-ES1 transition (red rectangle).
		    This pulse probes the initial population and drives spins into the non-resonant spin state (dark state). 
		    After a variable wait time $\tau$, a readout pulse of the same length follows (red rectangle) and probes the recovered population in the initial spin state. 
		    This sequence is repeated for different wait times $\tau$ and sample temperatures.
		    Error-bars of the individual data points are within their marker size.
		    The data for 4H-$\alpha$ is fitted with a double exponential function from which the decay rates are extracted.
		    \textbf{b)} Recovery signal for \SI{100}{\milli \second} pump pulse at \SI{1.3}{\kelvin}.
		    The double exponential behaviour vanishes and the data points follow a single exponential form.
		    For comparison the double exponential fit to the data from \textbf{a)} is shown as dashed line.
		    The resulting decay constant from the long pump pulse is in agreement with the value $\Gamma_1$ extracted from the short pulse.
		    \textbf{c)} Temperature dependence of the spin relaxation rate for 4H-$\alpha$ and 6H-$\alpha$ extracted from the data in panel a).
		    Circles correspond to the $\alpha$-site in 4H-SiC and triangles to 6H-SiC.
		    In 4H-SiC, two different time constants were extracted from the bi-exponential behaviour.
		    One with a temperature independent decay rate and one with a temperature dependent decay rate.
		    The data trace marked with 4H* (red circles) originates from a different setup (see text).
		    The vertical error bars are extracted from the fit to the data.
		    The error in temperature is given by its standard deviation, extracted from the values measured using a temperature sensor. 
		    \textbf{d)} Spin state depletion dependence on the applied magnetic field.
		    Depending on the strength of the magnetic field, the achievable spin depletion fraction varies and reaches a value of above \SI{80}{\percent} at \SI{0.49}{\tesla}. The data is fitted with a Lorentzian line shape.
		} 
		\label{fig:$T_1$temp}
	\end{figure*}
	\section{Electron spin lifetime}

	The electron spin lifetime of the ground state is an essential property for applications in quantum information and communication \cite{Ladd2010,Dolde2014,Kalb2017,Togan2010}.
    The spin lifetime needs to be sufficiently long for coherent spin manipulation, and hence sets a limit to the system's viability as qubit or quantum memory.
    Spin centres in wide band gap crystals are promising in this respect, as other defects, for example the nitrogen vacancy center in diamond, show exceptionally long spin lifetimes \cite{Astner2018}.
    
    We measure the spin relaxation for the 4H-$\alpha$ and 6H-$\alpha$ sites in the ground state at a magnetic field of \SI{0.49}{\tesla}.
    This field splits the spin states $\ket{GS1 \uparrow}$ and $\ket{GS1 \downarrow}$ by approximately \SI{12}{\giga\hertz}.
    To extract information about the $T_1$ spin lifetime, we employ an all optical spin depletion recovery sequence with transient detection.
    
    First, we initialize the system in a mixed state with a green laser pulse.
    Next, we apply a short laser pulse ($\leq \SI{1}{\milli\second}$) resonant with the GS1-ES1 transition, which results in pumping of the population into the opposite electron spin state. 
    During a variable wait time $\tau$, a fraction of the initial spin population recovers due to thermal relaxation.
    The recovered spin population is probed by a second resonant laser pulse.
    
    
    In \Cref{fig:$T_1$temp} a) we plot the time dependent spin population recovery for different temperatures in 4H-$\alpha$.
    The data is best described by a bi-exponential function with a fast ($\Gamma_0$) and slow ($\Gamma_1$) decay constant. 
    The fast relaxation shows no temperature dependence and has a constant rate $ \Gamma_0 $ of approximately \SI{100}{\per\second}.
	We show that this decay can be eliminated by increasing the pump pulse length to \SI{100}{\milli\second} $ \gg 1/\Gamma_0 $ (\Cref{fig:$T_1$temp} b)), hinting at a possible unknown shelving state within the system.
	The relaxation timescale of the mono-exponential recovery is in agreement with the slow decay rate $\Gamma_1$ determined using short initialization.
	We therefore identify the temperature-dependent relaxation rate $ \Gamma_1 $ as the spin-lattice relaxation rate and find the slowest measured rate to be \SI{0.04}{\per\second} at around \SI{100}{\milli\kelvin}  (\Cref{fig:$T_1$temp} c)), corresponding to a spin lifetime of \SI{25}{\second}.
	Increasing the temperature above \SI{1}{\kelvin} results in a spin lifetime of about \SI{1}{\second}.
	This temperature regime drastically reduces the cooling requirements, thus improving the prospects for practical implementation.
	This spin lifetime is fully sufficient for many of the intended applications.
	
		\begin{figure*}
		\centering
		\includegraphics[scale=1,page=3]{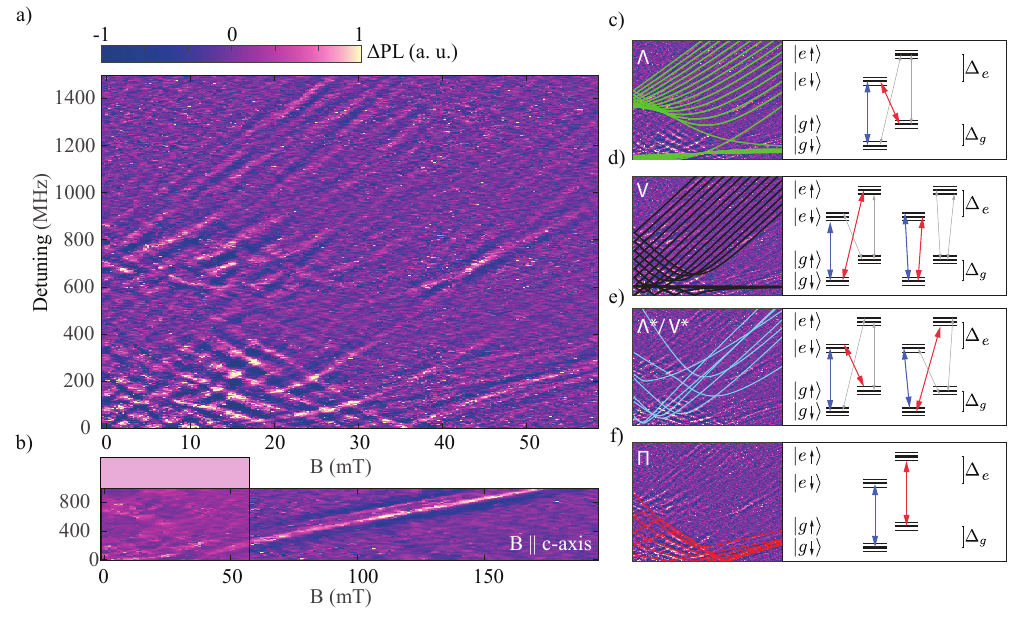}
		\caption{\textbf{Two-laser PLE spectroscopy data of 4H-$\mathrm{\alpha}$.}
			\textbf{a)} The PLE signal is recorded as function of the external magnetic field ranging from 0 to \SI{61}{\milli\tesla} and the two laser detuning (0 to \SI{1499}{\mega\hertz}) at a temperature $ <\SI{1}{\kelvin} $.
			The plot is rich in bright and dark features, which we use to refine the hyperfine parameters of the excited state. 
			\textbf{b)} Extended-range two laser spectroscopy: The bright features correspond to $\Pi$ transitions. 
			These transitions are visible when the Zeeman energy outweighs the hyperfine interaction.
			The shaded field corresponds to the data range in a).
			\textbf{c-f)} By diagonalizing the Hamiltonian we are able to plot the relevant transitions and can categorize the features into different groups:
			The coloured/grey arrows denote the two laser energies, connecting different states of the ground state manifolds and excited state manifolds.
			$\Delta_g$ and $\Delta_e$ denote the splitting of the different spin states by the electronic Zeeman effect.
			\textbf{c)} $ \Lambda $ transitions: The lasers address transitions from two different ground states to the same excited state, resulting in bright features.
			\textbf{d)} V transitions: The two laser beams are resonant with transitions associated with a common ground state.
			Population is driven to two different excited states.
			From there, the population can decay to other, undriven states, depleting the common ground state and resulting in dark lines in the signal.
			The two excited states can be within the same or in two different excited state manifolds, $ \ket{e\downarrow} $ and $ \ket{e\uparrow} $.
			\textbf{e)}Selected $\Lambda$* and V* transitions: $\Lambda$- and V-like systems are transition pairs that connect states in two different ground or excited state manifolds via two hyperfine states in a common excited or ground state electronic manifold, respectively. 
			\textbf{f)} $ \Pi $ (electron spin-conserving) transitions. Only transition pairs which connect states of the same nuclear spin at high field are shown. 
		}
		\label{fig:hyperfine}
	\end{figure*}
	
	We measured the spin-relaxation times of the V defects at higher temperatures using a different part of the 4H SiC wafer. These experiments were performed with a different experimental setup and time-resolved absorption of a pulsed resonant driving laser (see supplementary section \ref{suppleseclockin}). Between 1.8 K and 2.5 K, the spin-relaxation rate continues to increase with increasing temperature, reducing the relaxation lifetime to $\sim$\SI{100}{\micro\second} at \SI{2.4}{\kelvin} (see Fig. 2c). 

	As the recovery in 6H showed neither exponential nor bi-exponential behaviour, the measured 1/$e$ value was used to extract a value for the spin relaxation rate.
	Further details can be found in the supplementary materials section \ref{suplsec_tone}.
	For temperatures above \SI{1}{\kelvin} the 1/$e$ decay rate $\Gamma_1 $ shows a similar temperature scaling as 4H-$\alpha$.
	However, in the low temperature limit, the decay rate decreases far more slowly, and is more than an order of magnitude larger than in 4H.
	We measure the longest lifetime for the 6H-$\alpha$ site of $\sim$\SI{1}{\second} at \SI{150}{\milli\kelvin}.

Finally, we measured the bias field dependence of the spin depletion for both defects between \SIrange[]{0.03}{0.49}{\tesla}. As the magnetic field decreases, we observe a strong decrease in the spin contrast. The dependence of the spin depletion on the applied magnetic field for both 4H-$\alpha$ and 6H-$\alpha$ is shown in \Cref{fig:$T_1$temp} d).  This reduction of the spin depletion at low magnetic fields is likely due to spectral diffusion at magnetic fields where the electron Zeeman and hyperfine coupling are comparable \cite{Tissot2021,gilardoni21}. A Lorentzian fit to the data yields a linewidth of 0.6 GHz HWHM (half width at half maximum) in 6H-SiC, measured at \SI{0.15}{\kelvin},  and 0.8 GHz HWHM in 4H-SiC, measured at \SI{1.3}{\kelvin}. We interpret these values as the inhomogeneously broadenened single-spin linewidths. The field dependence is compatible with the assumption that the depletion is due to a change in the spin state of the electron, since all of the dominant optical transitions of the defect centre are known to lie within the depletion linewidth, as discussed in the following section. Changes in the relaxation rates could not be resolved in the photoluminescence measurements for temperatures below \SI{1.3}{\kelvin}. At  \SI{2}{\kelvin}, the absorption measurement reveals a strong magnetic field dependence of the relaxation lifetime, which decreases by more than a factor of three when decreasing the bias field from \SI{400}{\milli\tesla} to \SI{50}{\milli\tesla} (see supplementary section \ref{suplsec_magfield}).

\ifnum \bullets>0 

	\begin{itemize}
	
	    \item Motivation: $T_1$ never been seen, important for qubit
	    \item \textbf{Figure 3 a)}
	    \item Hole burning recovery sequence: green pulse, ir pump/readout pulse, wait time tau, read out pulse with IR, pulse length is 200 microsec
	    \item ref to \textbf{fig2 f} for pi transitions, resonant pump on GS1 - ES1 transition
	    \item at high field we see depletion during pump
	    \item IR power and duration is orders of magnitude less than in charge hole switching experiment, ref. supply. material to fidelity of burning vs. charge burning
	    \item dark time tau spins can recovery
	    \item probe shows amount of recovered spins
	    \item Transient detection of the recovery of the spin hole with different wait times tau between pump and readout pulse
	    \item Measurement for different temperatures, 0.1K up to 1.9K 
	    \item Description of the data with double exponential function, two decay constants, extracting lifetime from this data
	    \item \textbf{Figure 3 b)}
	    \item why 2 exponentials? try pumping longer than quick decay 
	    \item Varying the pulse sequence: intial preparation green pulse same as in a), pump pulse significantly longer, 200 micro sec --> 100 milli sec.,
	    \item Comparison of transient detection at at 1.3 K (add temp to plot!) for short and long pump pulse
	    \item Double exponential behaviour vanishes, signal recovery follows a single exponential, so with a long pump scheme, double exponential can be removed
	    \item \textbf{Figure 3 c)}
	    \item shows extracted decay constants from measurement for 4H
	    \item in 4H we see that gamma0 is constant for all measured temperatures
	    \item together with \textbf{fig 3 b} we conclude gamma 1 is spin $T_1$ gamma 0 unknown removable mechanism
	    \item in 4H gamma0 can be removed with the long pumping scheme and we end up with gamma1
	    \item in 4H gamma1 shows a temperature scaling
	    \item mention carmem data
	    \item present $T_1$ at 100mK 1.1 K and 2 K
	    \item measurement repeated for 6H alpha
	    \item neither exponential nor biexponential decay
	    \item in 6H we take the 1/e value as gamma1
	    \item temperature dependence, above 1K, similar behaviour to 4H gamma1
	    
	    \item \textbf{Figure 3 d)}
	    \item does $T_1$ have a B-field dependence???
	    \item we measure for different B fields
	    \item no difference in $T_1$ but pumping efficiency contrast changes drastically 
	    \item Depletion of the spin state by pumping it with an IR laser pulse
	    \item In suppl. mat. we show the pump sequence time trace
	    \item (send to suppl) We take the offset  value of each initial IR pump pulse from fitting the transient detection line shape at a temperature of 1.3K (4H). At 150 mK (6H) we take the average of the first three data points as a fit was not conclusive. This is then done for different magnetic fields B up to 480mT.
	    \item Fitting a Lorentzian line shape to the data points results in an optical line-width for the defect centers of: 6H - 0.6 GHz (HWHM +- 0.1GHz) and 4H - 0.8GHz (HWHM +- 0.3GHz).
	    \item linewidth is an upper bound, as we don't reach an equilibrium value
	    \item (suppl or fig) vertical error bars - 6H standard deviation of the three points
	    \item (suppl or fig) vertical error bars - 4H from fit
	    \item refer to \textbf{fig 2}
	    \item all comes together we know its spin pumping! and not nuclear! starting at 50 mT
	    
	\end{itemize}
	\fi

	\section{Hyperfine Structure}
	In order to better understand the observed shelving at high magnetic fields, further investigation into the magnetic field dependence of the optical transitions was conducted.
	Due to the inhomogeneously broadened transition \cite{Wolfowicz2020} between the optical ground and excited state of the ensemble, it is challenging to extract information on the underlying hyperfine structure of the defect with single laser spectroscopy.
	With a second laser frequency, it is possible to address two transitions between the ground and excited state hyperfine manifold simultaneously.
	This type of driving is only sensitive to the relative energy differences of the transitions, hence circumventing this limitation.
	We implement two-photon spectroscopy with a \SI{5}{\giga\hertz} electro-optical modulator that is driven with two RF frequencies to create two sets of optical sidebands: one at a constant detuning of \SI{4.5}{\giga\hertz} from the carrier, and a second set that can be swept from \SIrange{3}{4.5}{\giga\hertz}.
	Placing the carrier outside of the inhomogeneous lineshape, excitation by the carrier as well as by the mirror sidebands and higher order harmonics is efficiently suppressed, resulting in an effective single-sideband excitation spectrum, but without the need for quadrature modulation or sideband filtering (see supplementary materials section \ref{suplsec_twolaserscheme}). This method thereby allows to detect transition pairs separated by the frequency difference between the two sidebands. 
	
	The method samples from the edge of the inhomogeneous distribution, for which strain, impurities, or isotopic effects may modify the local crystal field. 
	This notwithstanding, no effects resulting from additional spins were resolved in the measurements, and there are no discernible shifts with respect to data taken near the peak of the distribution using singly-modulated excitation.
	
	In \Cref{fig:hyperfine} a) we show the observed PLE signal of the 4H-$\alpha$ site as function of the two-photon detuning and applied bias magentic field.
	6H-$\alpha$ exhibits very similar behaviour with the data set shown in the supplementary section \ref{suplsec_6htwoLaser}.
	We observe bright and dark features that can be mapped to different transitions by diagonalizing the Hamiltonian in \Cref{eq:sysham0}.
	We categorize the observed features into four different subsets:
	
	$ \Lambda $ transitions connect two different ground state levels via a common excited state.
    These transitions are visible as bright features, showing the ground state spin properties (\Cref{fig:hyperfine} c).
	Conversely, V-shaped transitions address two different excited-state levels from a common ground state (\Cref{fig:hyperfine} d). 
	As the involved excited states can decay to off-resonant ground state levels, the addressed ground state can be depleted, causing dark features in the observed fluorescence signal.
    The third category, displayed in \Cref{fig:hyperfine} e), are V-like  (V*)  or $\Lambda$-like ($\Lambda$*) systems, but with different hyperfine states in the common electronic manifold.
    These features reveal information about both the ground and excited state manifold.
	At high magnetic fields, the Zeeman term in the Hamiltonian exceeds the hyperfine coupling and effective spin conserving $\Pi$ transitions start to dominate the observed spectrum (\Cref{fig:hyperfine} b) and f)).
	From the slope of these transitions it is possible to determine the excited state g-factor: $\mathrm{g}=$\SI{2.18 \pm 0.02}{}.
	Efficient optical excitation in this system is spin conserving, but the excited states can decay via both spin-conserving and spin-flipping transitions.
	Spin flipping transitions emit orthogonally to the c-axis, and  can not be detected in our experiment.
	Thus only the spin conserving decays contribute to our fluorescence signal.
	The slope of these transitions as function of the magnetic field scale linearly with the difference between the ground and excited state electron g-factors (\Cref{fig:hyperfine} b).
	X type transitions, which are electron spin flipping but nuclear spin conserving, are not clearly identifiable in the measured spectra.
	We have therefore not included them in our analysis.
	The salient features in the two-photon spectroscopy allow a refined determination of the excited-state hyperfine parameters: $\mathrm{A_{xz} = \SI{75(4)}{\mega\hertz}, A_{zz}=\SI{-213(4)}{\mega\hertz}} $ (see \Cref{tab:table}).
	
	The large linewidths extracted from the field-dependent spin depletion measurements are in agreement with previously measured, single-defect linewidths \cite{Wolfowicz2020}, but in stark contrast with the narrow features observed in Fig. \ref{fig:hyperfine}.
	While the instantaneous linewidth of each defect may be narrow, laser illumination leads to a disturbance of the charge environment, and may cause spectral diffusion by shifting the resonance frequency via the Stark effect.
	This behaviour would have to be remedied for applications involving Purcell-enhanced photon emission, since the resulting inhomogeneous broadening will significantly reduce the performance of such a system   \cite{Chatzopoulos2019}.
	However, laser-induced dynamics would not be of concern for schemes using interaction-free readout and entanglement, since these methods rely on negligible excitation of the system \cite{Volz2011,Nemoto2014}.

\ifnum \bullets>0 

	\begin{itemize}
	    \item \textbf{Figure 2}
	    \item Motivation
	    \item Understand observed shelving at high field -> Important to understand level structure
	    \item also improtant for qubit building
	    \item still unresovled / partially resolved parameters in Hamiltonian
	    \item two laser spectroscopy (Method see suppl)
	    \item \textbf{FIG2 a} start at low field -> many many features
	    \item \textbf{Fig 2 b} at high field less but stronger
	    \item it is possible to assign features to transition families ref to \ref{eq:sysham} from GS to ES
	    \item present lambda v x pi style things
	    \item at high field pi is dominant refer back to \textbf{FIG1 d}
	    \item Salient features in the two-photon spectroscopy allow a refined determination of the excited-state g-factor and hyperfine parameters
	    \item present new values by michi
	    \item similar for 6H see suppl
	\end{itemize}
	\fi
	\ifnum \bullets>0 
	-Ground state features in low field, literature constants to fix I vs B
	
	-high-field dependence to fix g-g
	
	-combined yields g factor of excited state gZES 2.1908 
	
	-feature width of $\sim$\SI{90}{\mega\hertz} FWHM.
	
	-low-field V-type and X-type features to fix excited state hyperfine parameters
	
	-best match to observed features found with /. aXZ -> -75 /. aXX -> 0 /.aYY -> 0 /. aZZ -> 214
	
	- 
	
	-clearly identifiable features selected and show from many possible transitions.
	\fi

    \section{Summary and Conclusions}
    
The low-temperature measurements of the vanadium $\alpha$ sites in 4H-SiC and 6H-SiC presented here further underscore the potential of these defects for applications in quantum photonics. The long spin relaxation lifetime forms a strong basis for quantum memories and qubits, and provides ample time for entanglement distribution at the intercity scale. Nonetheless, several effects observed herein require further investigation.
Although the semi-insulating 4H-SiC and 6H-SiC samples are specified with the same background and vanadium dopant concentration, the charge state dynamics of the 4H- and 6H-$\alpha$ sites are significantly different.
Hole burning was observed in 4H, while the 6H defect did not present any bleaching. Furthermore, the single-exponential spin relaxation of the 4H site following enhanced initialization with long optical pulses was not reproduced in 6H-$\alpha$. The origin of these effects remains to be identified. Our measurements also indicate that the $\beta$ defects are most likely not suitable for such applications, since their spin relaxation lifetimes are below \SI{1}{\milli\second}, as described in the supplementary information. 

The high spin contrast can be expected to improve even further when addressing a single defect, given the limitations of fiber illumination of a thick crystal slab. Together with the detailed understanding of the spin structure gained from the two-photon spectroscopy measurements, this mechanism provides a first step towards the initialization of the spin into a pure state within the ground state manifold. Furthermore, the narrow features observed in the two-photon spectroscopy make the vanadium alpha defects highly attractive for use in spin-photon interfaces.

Given the prospects provided in this study, the vanadium defects merit intense further study regarding their quantum properties, particularly their coherence lifetimes and the characteristics of single defects. Like erbium in silicon, vanadium offers direct telecom interfacing between spins and photons, albeit with a much faster optical transition \cite{Dibos2015,Merkel2020}. The similarities with the silicon vacancy in diamond suggest that the spin coherence lifetime of vanadium will also be sufficient to enable the storage of quantum information for long-distance communication, providing a clear perspective towards high-performance quantum networks \cite{bhaskar2020,Reiserer2015,Kimble2008}.

	\begin{acknowledgments}
	We thank G. Burkard, B. Tissot and A. Gali for fruitful discussions.
	This work was funded from the European Union's Horizon 2020 Research and Innovation Program under grant agreement No. 862721 (QuanTELCO).
	N.T.S., I.G.I. and J.U.H. acknowledge support from the Knut and Alice Wallenberg Foundation (grant No. KAW 2018.0071).
	J.U.H also acknowledges support from Swedish Research Council under VR grant No. 2020-05444.

	T.A. and P.K. contributed equally to this work. 
	\end{acknowledgments}
	
	
	\bibliography{sicref}
	
\end{document}